\documentclass[reprint,amsmath,amssymb,aps,twocolumn,superscriptaddress,showpacs]{revtex4-1}
\usepackage{graphicx}
\usepackage{mathrsfs}
\usepackage{bm}
\usepackage{amsmath}
\usepackage{dcolumn}
\usepackage{epstopdf}
\usepackage{dsfont}
\usepackage{amssymb}
\usepackage{tabularx}
\usepackage{array}
\usepackage{float}
\usepackage{color}
\usepackage{epstopdf}
\usepackage{mathrsfs}
\makeatletter
\renewcommand*{\@fnsymbol}[1]{\ensuremath{\ifcase#1\or * \else * \fi}}

\makeatother
\usepackage[colorlinks=true, letterpaper=true, pdfstartview=FitV, linkcolor=blue, citecolor=blue, urlcolor=blue]{hyperref}
\begin{document}
\title{Two-dimensional antiferromagnetic semiconductor T'-MoTeI from first principles}

\makeatletter
\renewcommand*{\@fnsymbol}[1]{\ensuremath{\ifcase#1\or *\or \dag\or \dag \or
    \mathsection\or \mathparagraph\or \|\or **\or \dag\dag
    \or \dag\dag \else\@ctrerr\fi}}
\makeatother

\author{Michang Zhang}
\author{Fei Li}
\author{Yulu Ren}
\author{Tengfei Hu}
\author{Wenhui Wan}
\author{Yong Liu}
\author{Yanfeng Ge}
\email{yfge@ysu.edu.cn}\affiliation{State Key Laboratory of Metastable Materials Science and Technology \& Key Laboratory for Microstructural Material Physics of Hebei Province, School of Science, Yanshan University, Qinhuangdao, 066004, China}

\date{\today}

\begin{abstract}
   Two-dimensional intrinsic antiferromagnetic semiconductors are expected to stand out in the spintronic field. The present work finds the monolayer T'-MoTeI is intrinsically an antiferromagnetic semiconductor by using first-principles calculation. Firstly, the dimerized distortion of the Mo atoms causes T'-MoTeI to have dynamic stability, which is different from the small imaginary frequency in the phonon spectrum of T-MoTeI. Secondly, T'-MoTeI is an indirect-bandgap semiconductor with 1.35 eV. Finally, in the systematic study of strain effects, there are significant changes in the electronic structure as well as the bandgap, but the antiferromagnetic ground state is not affected. Monte Carlo simulations predict that the N\'{e}el temperature of T'-MoTeI is 95 K. The results suggest that the monolayer T'-MoTeI can be a potential candidate for spintronics applications.
\end{abstract}

\maketitle

\section{\label{sec:level1}INTRODUCTION}

Since the initial rush of research on graphene, two-dimensional (2D) materials have drawn wide attention due to their novel physical properties~\cite{KSNovoselov2004,KSNovoselov2005,AHCastroNeto2009,YMa2012,SLebegue2009}. Magnetically-ordered 2D crystals have brought infinite possibilities to development of new devices and discovery of physical phenomena. 2D magnetic semiconductors are of great potential for applications in low-dimensional spintronic devices because of their magnetic and semiconducting properties~\cite{MAshton2009,LZhou2015}. Many 2D magnetic materials have been predicted so far, and some have been successfully prepared in experiments~\cite{LDebbichi2016,HLiu2014,GRahman2017,VNicolosi2013,TJWilliams2015}. Because antiferromagnetic (AFM) semiconductors have stability under magnetic-field perturbations and large magneto-transport effects, 2D AFM semiconductors boast tremendous advantages in carrier injection, detection, sensors, and magnetic storage ~\cite{CGong2019}, but only a few 2D AFM semiconductors, such as Fe$_2$Cl$_3$I$_3$, FePS$_3$, and CoS$_2$, have been reported so far~\cite{LLiu2018,SChabungbam2017,YZhang2017,XXu2020,FKargar2020,ZZhang2020}. Further research on 2D intrinsic AFM semiconductors is necessary to promote practical applications.

In analysis of physical properties and applications of 2D materials, controlling the material structure is an effective method since two polymorphic structures may demonstrate drastically different properties despite the same chemical formula~\cite{MKan2014}. For instance, layered MoS$_2$, a famous transition-metal dichalcogenide (TMDC), has been extensively studied because of its distinctive electronic, optical, and mechanical properties ~\cite{QHWang2012}. Monolayer MoS$_2$ has three common structures, namely H-MoS$_2$, T-MoS$_2$ and T'-MoS$_2$~\cite{MAPy1983,FWypych1998,CAtaca2012,EBenavente2002}. H- and T-MoS$_2$ structures are semiconductor and metal, respectively. An electron crystallography study observed on the surface of restacked MoS$_2$ a new superstructure characterized by the formation of zigzag chains, which is called the distorted tetragonal MoS$_2$ (T'-MoS$_2$)~\cite{YHLee2012,ASingh2015,SNShirodkar2014,XRQin1991,MHWhangbo1992}. In addition to extensive research on pure TMDCs, recent advances in experimental techniques have allowed researchers to combine properties of different TMDCs in a single polar material, namely the Janus structure. The top layer of the monolayer MoTe$_2$ has been experimentally replaced by S atoms, namely MoSTe~\cite{HJin2018,ZWang2018}. Compared with the H-MoSTe structure, T'-MoSTe exhibits more Raman-active modes (a total of 15 modes). The T'-MoSTe structure makes the Mo-Mo bonds shorter and thus significantly enhances the in-plane stiffness. This structure turns into a piezoelectric material when out-of-plane symmetry is broken~\cite{MYagmurcukardes2019}. The obtainment of this monolayer compound has inspired many researchers to find a way to tune the physical properties of monolayer TMDCs. Synthesis of many polymorphic structures has been possible with the advancements of experimental techniques~\cite{THu2018}.

If a halogen element with more electrons is introduced into pure TMDCs to form a Janus structure, it will affect the occupation of electrons in the Mo-d orbital, resulting in magnetism regulated by structural distortion and strains. In the present work, it is found that the monolayer T'-MoTeI is an antiferromagnetic semiconductor. After dimerized distortion, T'-MoTeI is observed and characterized by the formation of zigzag chains. The formation energy of T'-MoTeI is lower than that of T-MoTeI. The calculated phonon spectrum of T'-MoTeI without imaginary frequency shows the dynamic stability of T'-MoTeI. In addition, the projected band structure is obtained to analyze the composition near the Fermi level. The properties of antiferromagnetic semiconductors are stable under biaxial strains, but the band gap value changes. Furthermore, we found that strain-free T'-MoTeI exhibits a T$_N$ at 95 K, which can be improved to 210 K by the appropriate strain condition.


\section{\label{sec:level} METHODS OF COMPUTATIONAL}

First-principles calculations were performed based on the density functional theory (DFT)~\cite{WKohn1965,PHohenberg1964} with the projector-augmented wave (PAW)~\cite{GKresse1995} method in the Vienna \emph{ab}-\emph{initio} Simulation Package (VASP)~\cite{GKresse1993,GKresse1996,JFurthmller1996}. The structural optimization and convergence tests were adopted with the Perdew-Burke-Ernzerhof (PBE)~\cite{JPPerdew1996} functional, and the electronic structures were given by the Heyd-Scuseria-Ernzerhof (HSE06) functional~\cite{TBjrkmanl2011}. The force and total energy convergence criteria were set at 0.1 meV/{\AA} and 10$^{-8}$ eV for all the calculations. The plane-wave cutoff energy, width of smearing, number of k points in reciprocal space were set to 550 eV, 0.05 eV, $9\times9\times1$, respectively. A 20{\AA} vacuum layer was applied between two nearest slabs to avoid the interlayer interactions. The $4\times4\times1$ supercell was applied to perform phonon calculations~\cite{ATogo2008}. In order to study the magnetic coupling between the magnetic moments on Mo atoms, we chose a supercell of $2\times2\times1$ in T'-MoTeI structure.

\section{\label{sec:level1} RESULTS and DISCUSSION}

\begin{figure}[tbp!]
\centerline{\includegraphics[width=0.45\textwidth]{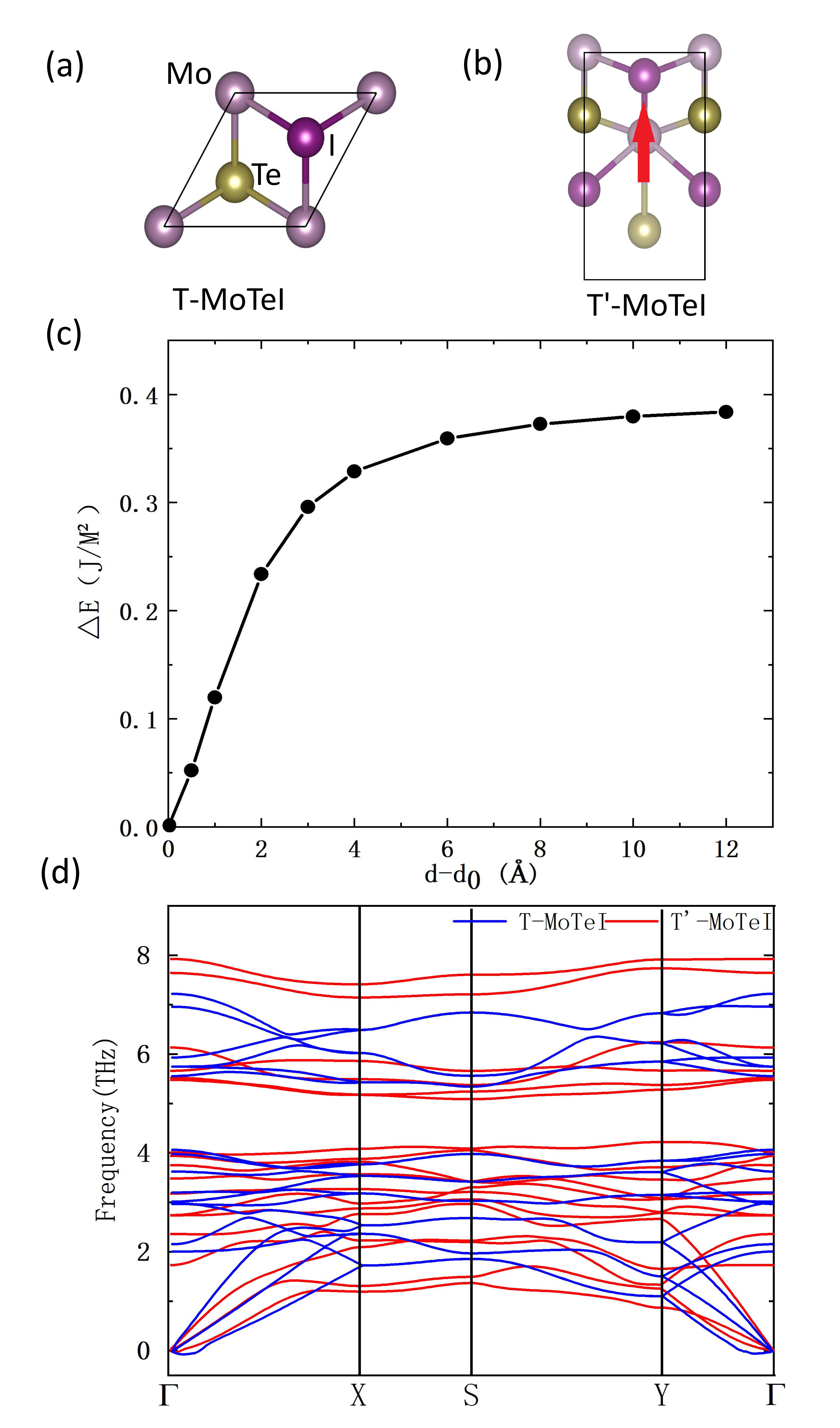}}
\caption{The top views of monolayer T-MoTeI and T'-MoTeI structures are (a) and (b) respectively. The Mo, Te and I atoms are in light gray, dark yellow and purple. (c) Increase in energy as a function of distance (relative to the equilibrium separation) between two half crystals of T'-MoTeI. (d) Phonon band structure of monolayer T'-MoTeI and T-MoTeI.
\label{fig:1}}
\end{figure}

In the T-MoTeI structure, the Mo atoms are octahedral coordinated with the nearby three Te and I atoms, resulting in ABC stacking with P3m1 space group symmetry [Fig.~\ref{fig:1}(a)]. The equilibrium lattice constant is 3.9{\AA} after high-precision structural optimization.
After the triangular lattice was changed into a rectangular primitive cell and the Mo atom was moved away from its original position, T'-MoTeI, as a new structure, was observed and characterized by the formation of zigzag chains as the same as that in T'-TMDCs [Fig.~\ref{fig:1}(b)]. T'-MoTeI is a charge density wave state as a result of the Piers phase transition from T phase~\cite{GEda2012,YGuo2015,GGao2015}.
In order to identify their preferred magnetic ground state, a ferromagnetic (FM) and three antiferromagnetic (AFM) configurations were constructed (see Appendix).
And the ground state is AFM$_2$ configuration, when FM and NM energies are much larger than three AFM states.

Comparison of energy of different magnetic configurations shows that AFM$_2$ is the most stable (see Appendix) and the magnetic moments are mostly localized on the magnetic centers of Mo (0.8 $\mu_B$) atoms, much more than those on Te (0.01 $\mu_B$) or I (0.01 $\mu_B$) atoms. It is worth noting that the ground state energy of T'-MoTeI is 1.98 eV, which is lower than that of T-MoTeI, and the following work mainly studies the ground state AFM$_2$ of T'-MoTeI. Then, we estimate the ease of exfoliation by the so-called cleavage energy, which is defined as the energy required to separate the crystal into two halves along the gap between two T'-MoTeI layers.
As shown in [Fig.~\ref{fig:1}(c)], the cleavage energy of T'-MoTeI is 0.37 J/m$^{2}$, close to those in other 2D materials, which indicates the feasibility of experimentally preparing T'-MoTeI. The phonon spectra also show that the distortion of dimerization makes T'-MoTeI dynamically stable, which is different from the T-MoTeI which has imaginary phonon frequencies [Fig.~\ref{fig:1}(d)].


\begin{figure*}[tbp!]
\centerline{\includegraphics[width=1\textwidth]{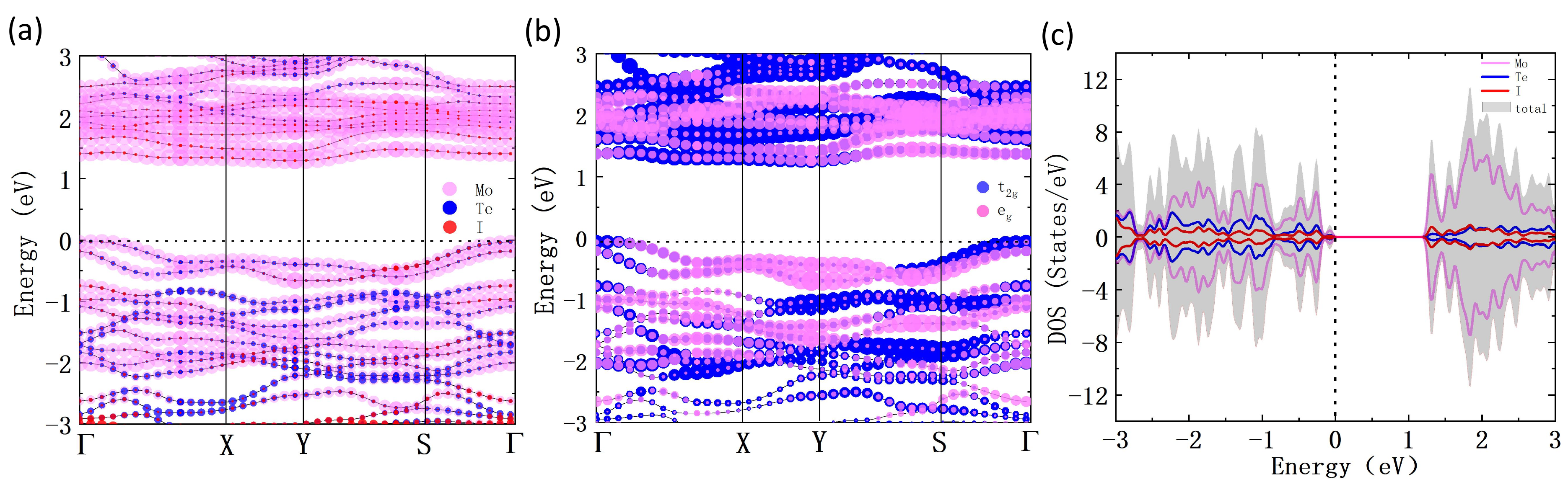}}
\caption{(a) Projected electronic band structures for monolayer T'-MoTeI. Mo, Te and I are denoted as pink, blue and red respectively. (b) Projected band structures for d orbits of Mo. (c) Total density of states (DOS) and project density of states (pDOS) of T'-MoTeI. In the DOS plot, the upper panel represents for the spin-up components and the lower panel for the spin-down components. The Fermi level is set to zero.
\label{fig:2}}
\end{figure*}

The spin projected band structure of monolayer T'-MoTeI are shown in [Fig.~\ref{fig:2}(a)]. The conduction band minimum (CBM) is located between Y and X points. The valence band maximum (VBM) is located between $\Gamma$ and X points. An indirect band gap of 1.35 eV is formed between CBM and VBM, smaller than the bandgap of T-MoTeI (1.80 eV), and indicates the intrinsic antiferromagnetic semiconductor of T'-MoTeI. The projected band structures and density of states (DOS) also shows that VBM and CBM are mainly contributed by Mo atoms when the contribution of Te-p and I-p orbitals are small around the bandgap [Fig.~\ref{fig:2}(b)]. Due to the distorted octahedral [Mo$_1$Te$_3$I$_3$] of T' structure, the Mo-d orbitals are roughly divided into upper e$_g$ (d$_{x^{2}-y^{2}}$, d$_{z^{2}}$) and lower t$_{2g}$ (d$_{xy}$, d$_{xz}$, d$_{yz}$) orbitals by the crystal-field effect. In both VBM and CBM, t$_{2g}$ contributes more energy than e$_{g}$. However, due to the influence of distortion, the e$_g$ and t$_{2g}$ orbitals are non-degenerate. Such a distortion further splits the d orbitals into five subgroups, as revealed by the orbital-resolved band structures. The orbitals are sequenced as follows by energy from low to high: d$_{xy}$, d$_{xz}$, d$_{yz}$, d$_{x^{2}-y^{2}}$, and d$_{z^{2}}$. Thus, for Mo$^{3+}$ in T'-MoTeI, there are three d electrons: two electrons completely fill the lowest d$_{xy}$ orbital, one electron half-occupies the d$_{xz}$ orbital. This results in one local magnetic moment for one Mo atom in T'-MoTeI.

Due to the different behaviors of d$_{Mo-Mo}$ and d$_{Mo-Te}$ under the strains, the mechanism of the AFM ground state and the strain effect on the magnetism can be understood by the competition between two different exchange interactions. On the one hand, the electrons occupying the t$_{2g}$ level have direct interaction between the nearest neighbor Mo atoms, which leads to the AFM arrangement.
 On the other hand, according to the Goodenough-Kanamori-Anderson (GKA) rules~\cite{JKanamori1959,JBGoodenough1995}, Mo-Te-Mo bond angle of 69$^{\circ}$ in T'-MoTeI is smaller than 80$^{\circ}$, the indirect interaction of the compound contributes to AFM~\cite{SLebernegg2013,ShijingZhang2009}. Further calculation shows that both negative direct and indirect interactions prove the AFM ground state of T'-MoteI.

\begin{figure*}[tbp!]
\centerline{\includegraphics[width=0.7\textwidth]{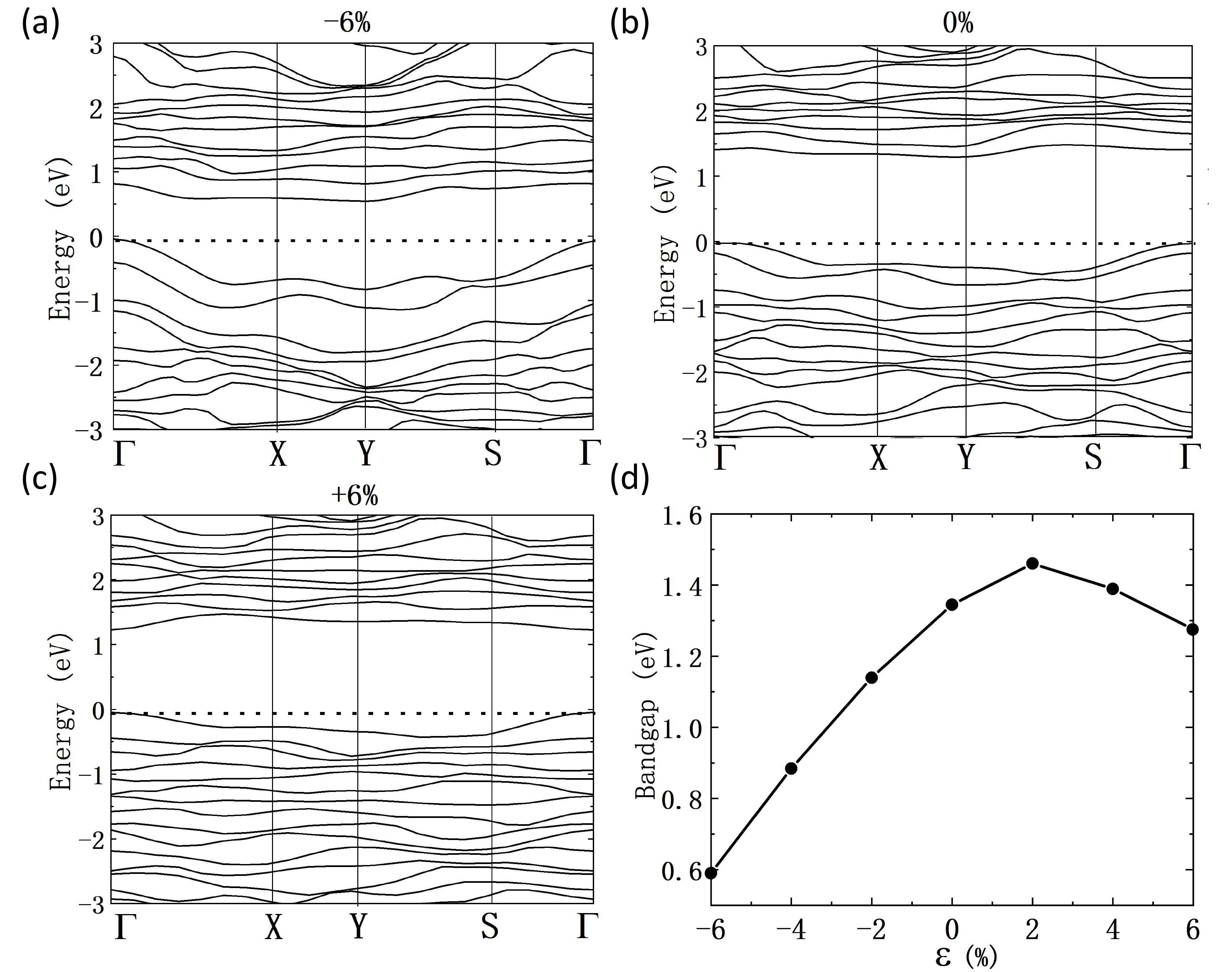}}
\caption{(a) PBE electronic band structure of monolayer T'-MoTeI at 6\%, (b) 0\%, (c) -6\% strains. (d) The band gap under different strains from 6\% to -6\%.
\label{fig:3}}
\end{figure*}

To explore the magnetic phase transitions in T-MoTeI and T'-MoTeI monolayers, a series of in-plane biaxial strains are applied. The value of strains applied are described by the change of the lattice parameter ${\varepsilon} = 100\%\times(a-a_0)/a_0= 100\%\times(b-b_0)/b_0$, where $a_0$ ($b_0$) and $a$ ($b$) are the lattice constants of unstrained and strained monolayer T'-MoTeI, respectively. The negative value denotes compressive strains and the positive value for tensile strains, and the range of strain is from -6\% to 6\%. As shown in [Fig.~\ref{fig:3}(a)], under a compressive strain of -6\%, the CBM locates at Y-point and VBM locates at $\Gamma$-point. While, both CBM and VBM at $\Gamma$-point form a direct band gap under a tensile strain of 6\% as shown in [Fig.~\ref{fig:3}(c)]. The band gap of T'-MoTeI can be tuned by as much as 50\% by applying strains [Fig.~\ref{fig:3}(d)]. Under tensile strains, the bandgap rises to the maximum at a 2\% strain and then decreases. For compressive strains, the bandgap decreases sharply.

\begin{figure}[tbp!]
\centerline{\includegraphics[width=0.5\textwidth]{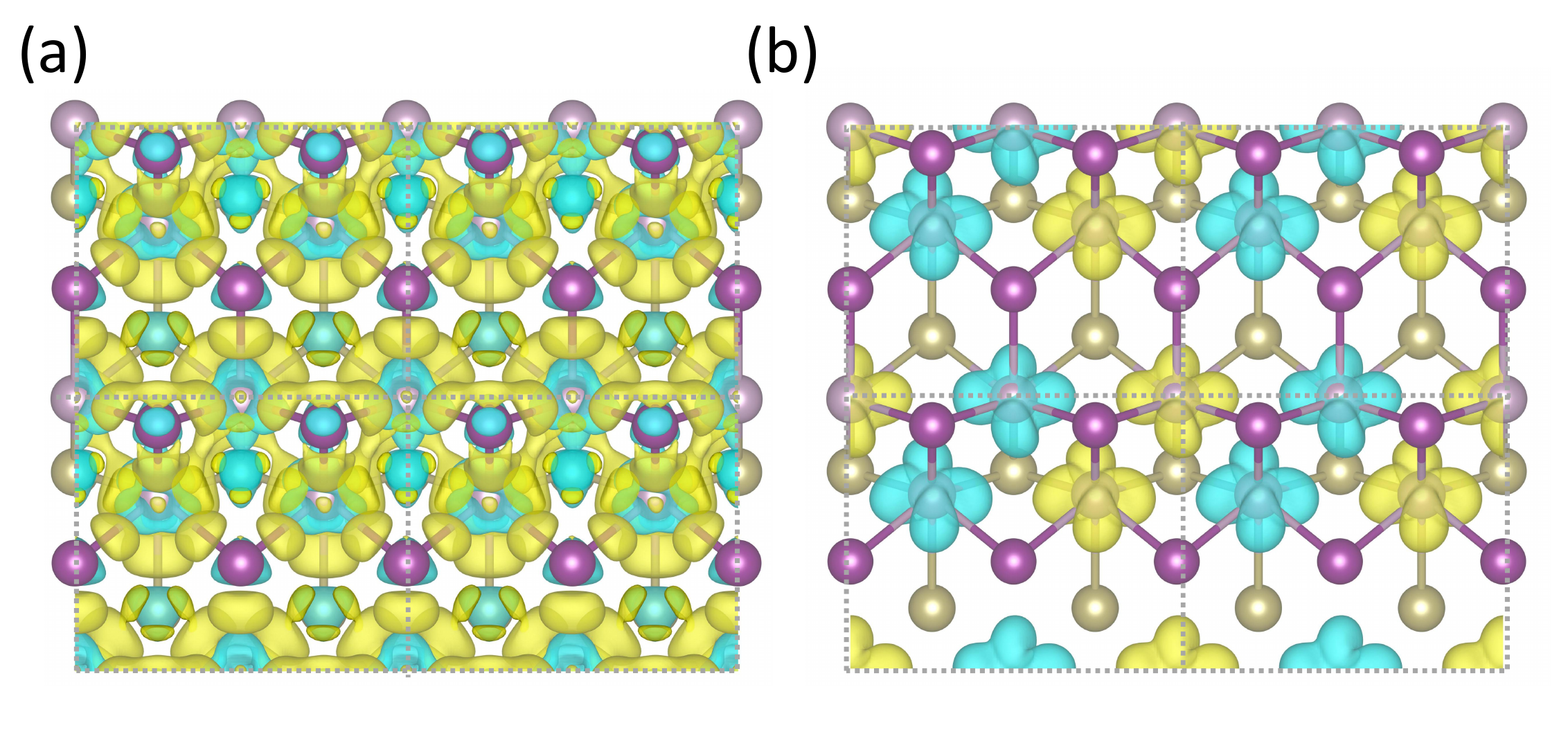}}
\caption{(a) Differential charge density of monolayer T'-MoTeI. (b) Isosurface of spin density with an isovalue.
\label{fig:4}}
\end{figure}

[Fig.~\ref{fig:4}(a)] shows the differential charge density of monolayer T'-MoTeI. It is defined as the difference between the charge density at the bonding point and the atomic charge density at the corresponding point. The yellow and blue regions represent charge accumulation and charge depletion, respectively. Significant charge redistributions are observed in Mo, Te and I atoms, where the Mo atoms lose electrons, while Te and I atoms gain electrons. This proves that the Mo element shows positive valence, but Te and I elements show negative valence. The spin densities for T'-MoTeI are shown in [Fig.~\ref{fig:4}(b)], where it can be observed that the spin-polarization mainly comes from Mo atoms, while the Te and I atoms are very small, which is consistent with the result of magnetic moment analysis.


\begin{figure}[tbp!]
\centerline{\includegraphics[width=0.5\textwidth]{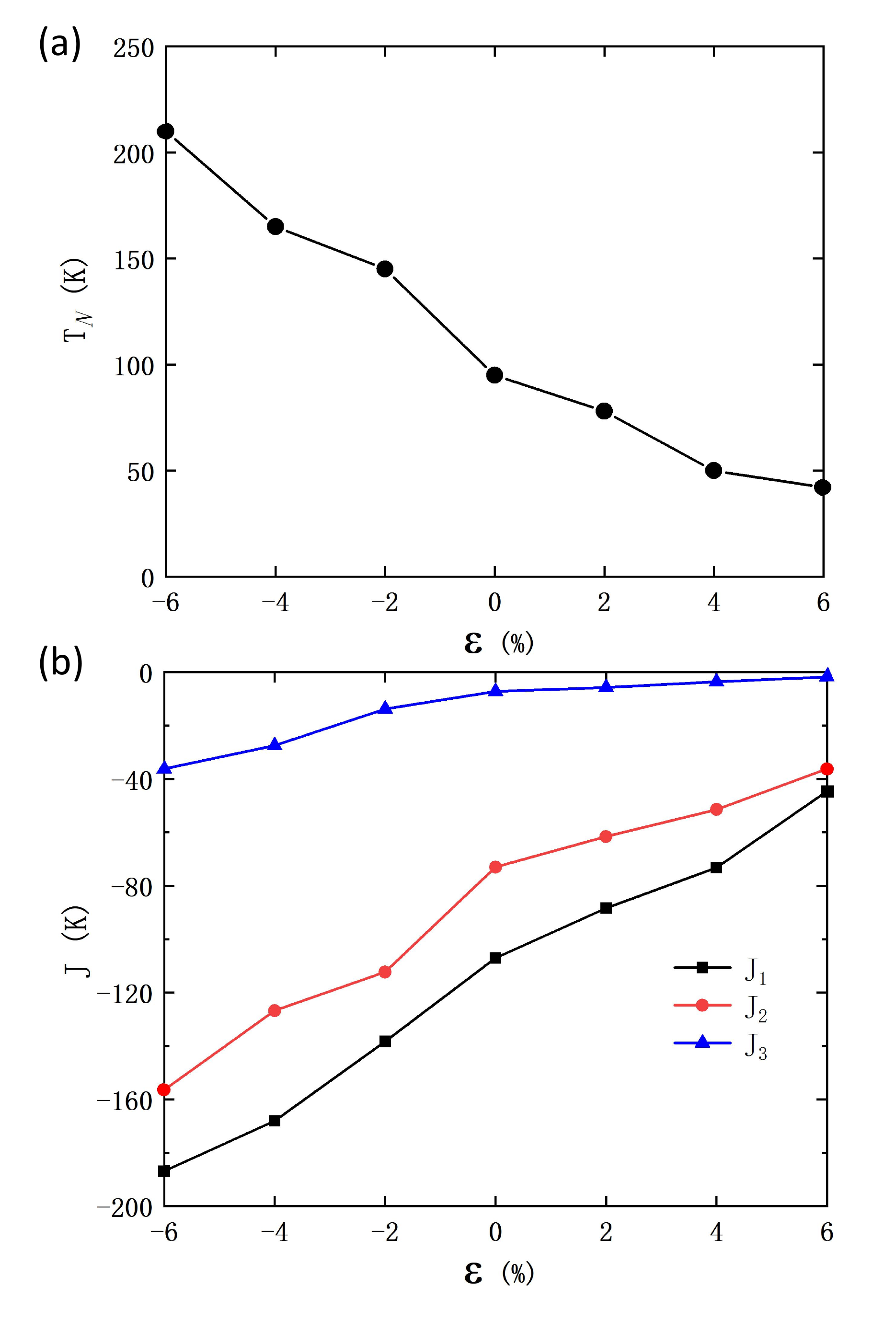}}
\caption{(a) The temperature of monolayer T'-MoTeI under different strains. (b) The interaction between the nearest neighbor Mo atoms (${J}_1$), the next nearest neighbor Mo atoms (${J}_2$) and the last nearest neighbor Mo atoms (${J}_3$) under biaxial strains. The energy units are in Kelvin.
\label{fig:5}}
\end{figure}

The N\'{e}el temperature (T$_N$) is a key parameter in applications of AFM materials to spintronic devices.
To explore the mechanism of magnetic phase transition, it is necessary to understand changes in magnetism with the temperature. We use the Ising model to characterize the magnetic coupling in monolayer T'-MoTeI. T$_N$ under different strains can be estimated by using MC simulations, as shown in [Fig.~\ref{fig:5}(a)]. The model Hamiltonian is written as
\begin{equation}
H =  - \sum\limits_{ < ij > } {{J_{ij}}{S_i}{S_j}},
\end{equation}
where J$_{ij}$ represents the exchange interactions of overall neighbor Mo-Mo pairs. Here we only consider the interaction between the nearest neighbor Mo atom (J$_1$), the next nearest neighbor Mo atom (J$_2$) and the third nearest neighbor Mo atom (J$_3$), respectively. In order to calculate the exchange interactions, the total energy of monolayer T'-MoTeI with FM, AFM$_1$, AFM$_2$ and AFM$_3$ configurations (see the Appendix) can be regard as
\begin{equation}{E_{FM}} = {E_0} - 2{J_1}{S^2} - 2{J_2}{S^2} - 2{J_3}{S^2},\end{equation}
\begin{equation}{E_{AFM1}} = {E_0} - 2{J_1}{S^2} - 2{J_2}{S^2} + 2{J_3}{S^2},\end{equation}
\begin{equation}{E_{AFM2}} = {E_0} + 2{J_2}{S^2},\end{equation}
\begin{equation}{E_{AFM3}} = {E_0} + 2{J_1}{S^2} - 2{J_2}{S^2} + 2{J_3}{S^2},\end{equation}
where $\emph{E}_0$ represents the energy of nonmagnetic T'-MoTeI monolayer. MC simulations reveal that monolayer strain-free T'-MoTeI has a critical temperature about 95 K. The strain-dependent T$_N$ mainly originated from the obvious strain effect on the exchange interactions.
As shown in [Fig.~\ref{fig:5}(b)], the J$_1$, J$_2$ and J$_3$ changes monotonously with strain.
According to the above equations about total energy, since J$_1$, J$_2$ and J$_3$ are both negative, the strain-free ground state is the magnetic configuration of AFM$_2$. It is found that the d$_{Mo-Mo}$ changes from 2.8{\AA} to 3.3{\AA} between -6\% and +6\% phase transition, while d$_{Mo-Te}$ is almost unchanged. This indicates that the changes of AFM strength in strains mainly come from the direct interaction between Mo atoms. Under the compressive strains, significantly reduced distance between the Mo atoms enhances the direct exchange interaction. For the considered largest compressive strain ($\varepsilon$= -6\%), the J$_1$ and T$_N$ values were -16 meV and 210 K, respectively. While, with the increase of d$_{Mo-Mo}$ under the tensile strain, the corresponding J, and T$_N$ values decrease because of the weakened direct exchange interaction.


In conclusion, we studied the magnetic and electronic properties of T'-MoTeI monolayer using first-principles calculations. The phonon calculation indicates the structural stability of T'-MoTeI. Further studies on differential magnetic configurations and the critical temperature reveal that the monolayer T'-MoTeI has a N\'{e}el temperature above 95 K under strain-free condition. The present work also revealed the obvious impact of strains on the magnetic and electronic properties of T'-MoTeI. Our results show that the monolayer T'-MoTeI, as a new two-dimensional ferromagnetic semiconductor, will have great potential for nanoscale spintronic applications.

\begin{acknowledgments}
This work is supported by National Natural Science Foundation of China (No. 11904312 and 11904313), and the Natural Science Foundation of Hebei Province of China (No. A2020203027).
\end{acknowledgments}

\appendix*
\renewcommand\thefigure{A\arabic{figure}}
\setcounter{figure}{0}
\section{A}
\begin{figure}[H]
\centerline{\includegraphics[width=0.45\textwidth]{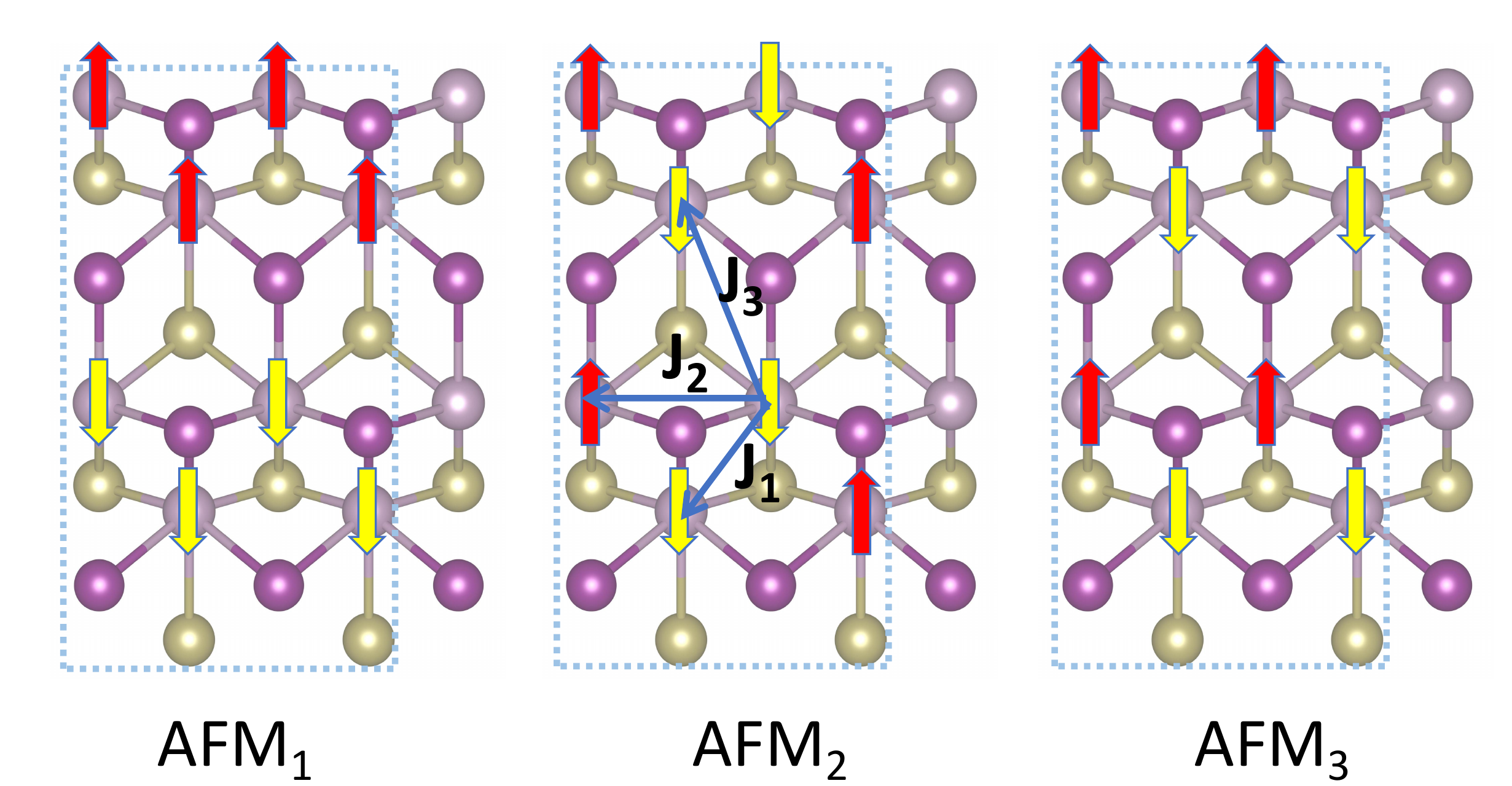}}
\caption{T'-MoTeI different arrangements of spins considered in this work.
\label{fig:S1}}
\end{figure}
\begin{figure}[H]
\centerline{\includegraphics[width=0.5\textwidth]{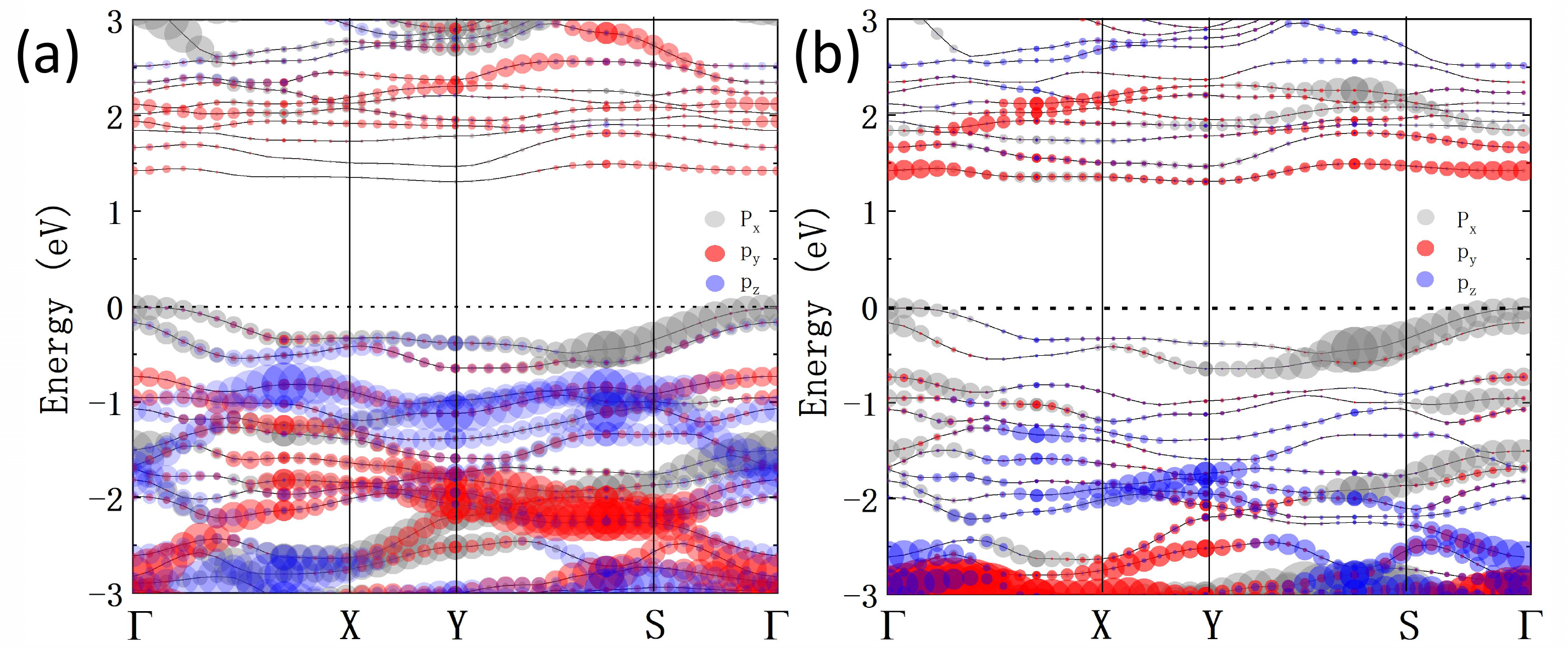}}
\caption{Projected band structures for p orbits of Te and I atoms.
\label{fig:S2}}
\end{figure}



\nocite{*}

\end{document}